\documentclass[a4paper,12pt]{article}
\usepackage{epsfig}
\usepackage{subfigure}
\usepackage{amssymb, graphics}
\usepackage{pslatex}
\usepackage{graphicx}
\usepackage{bbm}
\usepackage{latexsym}
\usepackage{amsfonts}
\usepackage{amsmath}
\usepackage[title,titletoc]{appendix}
\usepackage[super]{nth}
\usepackage{cancel}

\newcommand{\mathsym}[1]{{}}

\newcommand{\qed}{\nobreak \ifvmode \relax \else \ifdim\lastskip<1.5em \hskip-\lastskip \hskip1.5em plus0em minus0.5em 
\fi \nobreak \vrule height0.75em width0.5em depth0.25em\fi}

\def\app#1#2{  \mathrel{    \setbox0=\hbox{$#1\sim$}    \setbox2=\hbox{      \rlap{\hbox{$#1\propto$}}     
 \lower1.1\ht0\box0    }    \raise0.25\ht2\box2  }}

\begin{document}

\begin{titlepage}
\begin{center}

\vskip 1cm

{\large \bf Quasidegeneracy of Majorana Neutrinos  \\
and the Origin of Large Leptonic Mixing}

\vskip 1cm

G.C. Branco$^{a,b,}$\footnote{E-mail: gbranco@tecnico.ulisboa.pt}, 
M.N. Rebelo$^{a,}$\footnote{E-mail: rebelo@tecnico.ulisboa.pt}
J.I. Silva-Marcos$^{a,}$\footnote{E-mail: juca@cftp.ist.utl.pt}
Daniel Wegman$^{a,}$\footnote{E-mail: wegman.daniel@gmail.com}

\vskip 0.07in

{\em 
$^a$Centro de F{\'\i}sica Te\'orica de Part{\'\i}culas - CFTP, \\
$^b$ Departamento de F\'{\i}sica,\\}
\vskip 0.5cm

{\it  Instituto Superior T\'ecnico - IST, Universidade de Lisboa - UL, }
\\
{\it
Avenida Rovisco Pais, 1049-001 Lisboa, Portugal}
\end{center}

\vskip 3cm

\begin{abstract}
We propose that the observed large leptonic mixing may just reflect a quasidegeneracy
of three Majorana neutrinos. The limit of exact degeneracy of Majorana neutrinos is 
not trivial, as leptonic mixing and even CP violation may occur. We conjecture that the 
smallness of $|U_{13}|$, when compared to the other elements of $U_{PMNS}$, may be
related to the fact that, in the limit of exact mass degeneracy, the leptonic mixing
matrix necessarily has a vanishing element. We show that the lifting of the mass degeneracy
can lead to the measured value of $|U_{13}|$ while at the same time accommodating the observed
solar and atmospheric mixing angles. In the scenario we consider for the breaking of the mass degeneracy
there is only one CP violating phase, already present in the limit of exact degeneracy, 
which upon the lifting of the degeneracy generates both Majorana and Dirac-type CP 
violation in the leptonic sector.  We analyse some of the correlations among physical
observables and point out that in most of the cases considered, the implied strength of 
leptonic Dirac-type CP violation is large enough to be detected in the next round of
experiments.

\end{abstract}

\end{titlepage}

\newpage

\section{Introduction}

The observed pattern of fermion masses and mixing continues being a major
puzzle in particle physics and the discovery of large leptonic mixing
rendered the question even more intriguing. A large number of models have
been suggested in the literature for providing an understanding of neutrino
masses and mixing. These models cover a large number of possibilities, going
from models with discrete abelian or non-abelian symmetries \cite{symmodels}
to the suggestion that in the neutrino sector, anarchy prevails \cite%
{anarchy}.

In this paper we conjecture that the observed large mixing in the lepton
sector may just reflect a Majorana character of neutrinos and
quasidegeneracy of neutrino masses. It is well known that, for Dirac
neutrinos, leptonic mixing can be rotated away in the limit of exact
neutrino mass degeneracy. For Majorana neutrinos, it has been pointed out
that leptonic mixing and even CP violation can occur in the limit of exact
neutrino mass degeneracy and in this limit, leptonic mixing is characterized
by two angles and one CP violating phase \cite{Branco:1998bw}. In this
limit, the leptonic unitarity triangles are collapsed in a line, since one
of the entries of the leptonic mixing vanishes, thus implying no Dirac-type
CP violation. However, the Majorana triangles do not all collapse into the
real or imaginary axis, thus implying \cite{AguilarSaavedra:2000vr} CP
violation of Majorana type. We identify the zero entry of the leptonic
mixing matrix with $U_{13}$ and show that a small perturbation around the
degenerate limit generates the observed neutrino mass differences as well as
leptonic mixing in agreement with experiment, including the recent
measurements of the smallest mixing angle, $\theta _{13}$, at reactor \cite%
{reactor}, and accelerator \cite{Abe:2013xua} neutrino experiments. In this
framework, one also finds a possible explanation for the smallness of $%
|U_{13}|$, compared to the other entries of the $U_{PMNS}$. This may just
reflect the fact that, in the exact degenerate limit of Majorana neutrinos,
one of the entries of $U_{PMNS}$ necessarily vanishes.

As soon as it became clear that the experimental evidence favoured a
nonvanishing $U_{13}$, many proposals \cite{aftera4} were put forward in the
literature analysing how small perturbations around various textures
obtained from symmetries, could accommodate a non-vanishing $U_{13}$ while
also correctly reproducing the data on the solar and atmospheric mixing
angles. The distinctive feature of our proposal is the fact that we start
from the non-trivial limit of exactly degenerate Majorana neutrinos.

This paper is organised as follows. In the next section, we study the limit
of exact degeneracy of three Majorana neutrinos, pointing out that in this
limit the Majorana mass matrix is proportional to a unitary matrix and
describing the implications for leptonic mixing and CP violation. In section
3, we study the lifting of the mass degeneracy with the generation of
neutrino mass differences and a non-vanishing $U_{13}$. We analyse in detail
the case where the unperturbed leptonic mixing is given by some of the most
popular Ans\"{a}tze, allowing for Majorana-type CP violation, with special
emphasis on the tribimaximal case \cite{Harrison:2002er}. We consider a
scenario for the breaking of the degeneracy, where there is only one CP
violating phase which, upon the lifting of the degeneracy, generates both
Majorana and Dirac-type CP violation. Finally, in section 4 we present our
conclusions.

\section{The Limit of Exact Degeneracy}

\subsection{The Majorana Neutrino Mass Matrix}

Without loss of generality, we choose to work in a weak basis (WB) where the
charged lepton mass matrix is diagonal, real and positive. We assume three
left-handed neutrinos and consider a Majorana mass term with the general
form: 
\begin{equation}
\mathcal{L}_{{\mathrm{mass}}}\ =\ -\ ({\nu }_{{_{L}}_{\alpha }})^{T}\
C^{-1}\ (M_{o})_{\alpha \beta \quad }{\nu }_{{_{L}}_{\beta }}\ +\mathrm{\
h.c.}  \label{eq1}
\end{equation}
where ${\nu }_{{_{L}}_{\alpha }}$ stand for the left-handed weak eigenstates
and $M_{o}$ is a $3\times 3$ symmetric complex mass matrix. Since in general 
$M_{o}$ is diagonalized by a unitary matrix $U_{o}$ through $U_{o}^{T}\
M_{o}\ U_{o}$ $=diag$ $(m_{\nu _{1}},m_{\nu _{2}},m_{\nu _{3}})$, it follows
that in the limit of exact neutrino mass degeneracy, $M_{o}$ can be written: 
\begin{equation}
M_{o}={\mu }\ S_{o}
\label{eq2}
\end{equation}
where ${\mu }$ is the common neutrino mass and $S_{o}=U_{o}^{\ast
}U_{o}^{\dagger }$. In the limit of exact degeneracy, a novel feature
arises, namely $M_{o}$ is proportional to the symmetric unitary matrix $%
S_{o} $. Under a WB transformation corresponding to a rephasing of both $\nu
_{_{L}}$ and the charged lepton fields, the neutrino mass matrix transforms
as: 
\begin{equation}
M_{o}\rightarrow L\ M_{o}\ L  \label{rot}
\end{equation}%
with $L\equiv diag(e^{i\varphi _{1}},e^{i\varphi _{2}},e^{i\varphi _{3}})$.
As a result, the individual phases of $M_{o}$ have no physical meaning, but
one can construct polynomials in $(M_{o})_{ij}$ which are rephasing
invariant \cite{Farzan:2006vj} such as $(M_{o}^{\ast })_{11}(M_{o}^{\ast
})_{22}(M_{o}^{\ast })_{12}^{2}$ or $(M_{o})_{11}(M_{o}^{\ast
})_{33}(M_{o})_{13}^{2}$. The fact that $S_{o}$ is symmetric and unitary
implies that in general $S_{o}$ can be parametrized by two angles and one
phase. In Ref. \cite{Branco:1998bw}, the limit of exact degeneracy for
Majorana neutrinos was analysed in some detail and it was shown that
leptonic mixing and even CP violation can occur in that limit. Leptonic
mixing can be rotated away if and only if there is CP invariance and all
neutrinos have the same CP parity \cite{Wolfenstein:1981rk}, \cite%
{Branco:1999fs}. Furthermore, it was also shown in Ref \cite{Branco:1998bw}
that in the case of different CP parities, the most general matrix $S_{o}$
can be parametrized in terms of two rotations with three-by-three orthogonal
matrices having only a two-by-two non diagonal block each, corresponding to
a single mixing angle, together with one diagonal matrix with one phase: 
\begin{equation}
S_{o}\ =\ \left( 
\begin{array}{ccc}
1 & 0 & 0 \\ 
0 & c_{\phi } & s_{\phi } \\ 
0 & s_{\phi } & -c_{\phi }%
\end{array}%
\right) \cdot \left( 
\begin{array}{ccc}
c_{\theta } & s_{\theta } & 0 \\ 
s_{\theta } & -c_{\theta } & 0 \\ 
0 & 0 & e^{i\alpha }%
\end{array}
\right) \cdot \left( 
\begin{array}{ccc}
1 & 0 & 0 \\ 
0 & c_{\phi } & s_{\phi } \\ 
0 & s_{\phi } & -c_{\phi }
\end{array}
\right)  \label{eq6}
\end{equation}
this equation is of the form: 
\begin{equation}
S_{o}\ =O_{23}(\phi )\ O_{12}(\theta )\ \left( 
\begin{array}{ccc}
1 & 0 & 0 \\ 
0 & 1 & 0 \\ 
0 & 0 & e^{i\alpha }
\end{array}
\right) \ O_{23}({\small \phi })  \label{eq7}
\end{equation}
with each orthogonal matrix $O_{ij}$ chosen to be symmetric. Using the fact
that $S_{o}=U_{o}^{\ast }U_{o}^{\dagger }$ one concludes that , in this limit,
the leptonic mixing matrix is given by: 
\begin{equation}
U_{o}\ =O_{23}\left( {\small \phi }\right) {\small \ }O_{12}\left( \frac{%
{\small \theta }}{{\small 2}}\right) \ \left( 
\begin{array}{ccc}
1 & 0 & 0 \\ 
0 & i & 0 \\ 
0 & 0 & e^{-i\frac{\alpha }{2}}
\end{array}
\right)  \label{eq8}
\end{equation}
up to an orthogonal rotation of the three degenerate neutrinos.

Given the Majorana character of neutrino masses, it is clear that even in
the limit of exact degeneracy with CP conservation, but with different
CP-parities, one cannot rotate $U_{o}$ away through a redefinition of the
neutrino fields. It should be emphasized that the leptonic mixing matrix
is only defined up to an orthogonal rotation of the three 
degenerate neutrinos. Indeed if $U_{o}$ diagonalizes $M_{o}$ so does 
$U_{o}O$, as it is evident from Eq. (\ref{eq2}) 
and the fact that $S_{o} =U_{o}^* U_{o}^\dagger$ . 
Without loss of generality one can eliminate the matrix O. 
It is important to notice that $U_{o}$ always has one zero
entry which in the above parametrization appears in the $(13)$ position.
This may be a hint that the limit of exact degeneracy is a good starting
point to perform a small perturbation around it, leading to the lifting of
the degeneracy and the generation of a non-zero $U_{e3}$. At this stage, it
should be noted that although the limit of exact degeneracy necessarily 
implies a zero entry in $U_{o}$ the location of the zero is not fixed. If we
had interchanged the r\^{o}les of $O_{23}$ and $O_{12}$ in Eq. (\ref{eq6}),
the zero entry would appear in the $(31)$ position. Our choice of Eq. (\ref
{eq6}) was dictated by the experimental fact that the leptonic mixing matrix
has a small entry in the $(13)$ position. It should be stressed that the identification of 
$U_{PMNS}$ with $U_{o} O$ can only be done  after the lifting of the degeneracy,  
which will be done in the sequel. The matrix $O$ will be fixed by the perturbation 
of $M_{o}$ leading to the lifting of the degeneracy. 
In the exact degenerate limit the individual elements of the matrix $U_{o} O$
have no physical meaning. But there are physical quantities which do have physical 
meaning even in the exact degenerate limit. These quantities are independent of the 
matrix $O$, depending only on combinations of the angles $\theta$, $\phi$ 
and the phase $\alpha$,  entering in the parametrisation of $U_o$  given in Eq. (\ref{eq6}).
An example of such a physical quantity, will be given in the next subsection, where 
we evaluate in the exact degenerate limit the strength of Majorana type CP violation, 
expressed in terms of the mixing angles $\theta$, $\phi$  and the phase $\alpha$.
Of course, this quantity does not depend on the matrix $O$.

It is easy to understand why a symmetric unitary $3 \times 3$ matrix, such
as $S_{o}$, can be parametrized by only two angles and one phase. On one
hand, there is the freedom of choice of WB given by Eq.~(\ref{rot}) on the
other hand for a general $3\times 3$ unitary matrix $U$, one can define an
asymmetry parameter, given by \cite{Branco:1990zq}: 
\begin{equation}
A_s \equiv
|U_{12}|^{2}-|U_{21}|^{2}=|U_{31}|^{2}-|U_{13}|^{2}=|U_{23}|^{2}-|U_{32}|^{2}
\end{equation}
In the case of a unitary symmetric matrix, one has $A_s =0$, which leads to
the loss of one parameter.

The parametrization of $S_{o}$ in terms of two rotations and one phase is
the most general one (apart from the unphysical complex phases which can be
rotated away as in Eq. (\ref{rot})). This can be seen by recalling that the
parametrization of a general unitary matrix through Euler angles involves
three orthogonal rotations, usually denoted by $O_{12}$, $O_{13}$, $O_{23}$.
The fact that $S_{o}$ is symmetric implies the loss of one parameter and, as
a result, only two orthogonal matrices are needed. The rotation matrix 
$O_{rs}$ that is left out, dictates the entry of $U_{o}$ that is zero to be, 
$(r,s)$ or $(s,r)$ depending on the order chosen for the other two orthogonal
matrices.

\subsection{{$S_{o}$} unitarity triangles, leptonic mixing and CP violation}

Since $S_{o}$ is a unitary matrix, one can consider $S_{o}$ unitarity
triangles, which are analogous to the ones \cite{AguilarSaavedra:2000vr}
encountered in the leptonic mixing matrix $U_{PMNS}$, but with a different
physical meaning. A unitarity triangle corresponding to orthogonality of the
two first columns of $S_{o}$ is displayed in Figure 1. 
\begin{figure}[h]
\begin{center}
\includegraphics[scale=1.3]{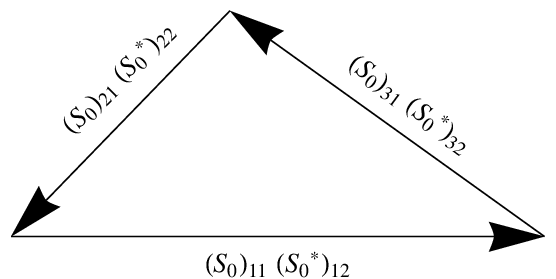}
\end{center}
\caption{Unitarity triangle built from the first two columns of $S_{0}$, for
a generic unitary matrix, assuming CP violation}
\label{figura1}
\end{figure}
Note that the orientation of the $S_{o}$ unitarity triangles rotates under
the rephasing of Eq.(\ref{rot}) and therefore it has no physical meaning.
However, the area of the $S_{o}$ triangles has got physical meaning, giving
a measure of the strength of Majorana-type CP violation in leptonic mixing
in the case of exact degeneracy. All $S_{o}$ unitarity triangles have the
same area $A$, which equals twice the absolute value of any of the rephasing
invariant quartets $Q_{s}$ of $S_{o}$: 
\begin{equation}
A=2|\mbox{Im}Q_{s}|=\frac{1}{2}\left\vert \cos \left( \theta \right) \sin
^{2}(\theta )\sin ^{2}(2\phi )\sin \left( \alpha \right) \right\vert
\label{aqs}
\end{equation}%
with $|Q_{s}|\equiv |(S_{o})_{ij}(S_{o})_{ik}^{\ast }(S_{o})_{lj}^{\ast
}(S_{o})_{lk}|$ with $i\neq l$, $j\neq k$. In the limit of exact degeneracy,
we have seen that the leptonic mixing matrix $U_{o}$ has a zero entry, which
implies that there is no Dirac-type CP violation and all the $U_{o}$
unitarity triangles collapse to lines. However, there is CP violation of the
Majorana-type, since the Majorana unitarity triangles for $U_{o}$ are in
general not collapsed along the real and imaginary axis \cite%
{AguilarSaavedra:2000vr}. Once the degeneracy is lifted the leptonic
unitarity triangles open up and Dirac-type CP violation is generated. In
Ref.~\cite{Branco:2008ai}, it was shown how to express, in this case, 
the full PMNS matrix, including the
strength of Dirac-type CP violation in terms of arguments of the six
independent rephasing invariant bilinears corresponding to the 
orientation of the sides of
Majorana-type unitarity triangles, thus showing that Dirac-type CP violation
in the leptonic sector with Majorana neutrinos, necessarily implies Majorana-type CP violation.

\section{Lifting the Degeneracy}

\subsection{Rationale and Strategy}

For definiteness and without loss of generality, we work in the weak basis
where the charged lepton mass matrix is diagonal real. As emphasized in the
previous section, in the case of exactly degenerate Majorana neutrinos,
mixing is meaningful and it can be parametrized by two angles and one phase. 

Several textures for the leptonic mixing matrix have been studied in the
literature, often in the context of family symmetries \cite{symmodels}. In
most of the proposed schemes, the pattern of leptonic mixing is predicted
but the spectrum of masses is not constrained by the symmetries. It is
therefore consistent to consider these schemes, together with the hypothesis
of quasidegeneracy of Majorana neutrinos. A different approach connecting
the leptonic mixing parameters with certain kinds of degeneracy of the
neutrino mass spectrum was followed in \cite{Hernandez:2013vya}.

Until recently one of the most favoured Ans\"{a}tze, from the experimental
point of view, seemed to be the tribimaximal mixing \cite{Harrison:2002er}
which has a zero in the $(13)$ entry. Other interesting textures which also
have a zero entry in this location \cite{Wang:2013wya} include the
democratic mixing \cite{Fritzsch:1995dj}, bimaximal mixing \cite%
{Barger:1998ta}, golden ratio mixings \cite{Kajiyama:2007gx}, \cite%
{Rodejohann:2008ir}, hexagonal mixing \cite{Albright:2010ap} and bidodeca
mixing \cite{Kim:2010zub}, \cite{Kim:2011vr}. Recent measurements of $\theta
_{13}$, the smallest of the mixing angles of $U_{PMNS}$ as given by the
standard parametrization \cite{Beringer:1900zz}, have established a non-zero
value for this angle \cite{Tortola:2012te}. In the standard parametrization,
$U_{PMNS}$ is given by: 
\begin{equation}
U_{PMNS}=\left( 
\begin{array}{ccc}
c_{12}c_{13} & s_{12}c_{13} & s_{13}e^{-i\delta } \\ 
-s_{12}c_{23}-c_{12}s_{23}s_{13}e^{i\delta } & \quad
c_{12}c_{23}-s_{12}s_{23}s_{13}e^{i\delta }\quad & s_{23}c_{13} \\ 
s_{12}s_{23}-c_{12}c_{23}s_{13}e^{i\delta } & 
-c_{12}s_{23}-s_{12}c_{23}s_{13}e^{i\delta } & c_{23}c_{13}%
\end{array}%
\right) \,\cdot P,  \label{pdg}
\end{equation}
where $c_{ij}\equiv \cos \left( \theta _{ij}\right) \ ,\ s_{ij}\equiv \sin
\left( \theta _{ij}\right) \ $, with all $\theta _{ij}$ in the first
quadrant, $\delta $ is a Dirac-type phase and $P=diag(1,e^{i\alpha
},e^{i\beta })$ with $\alpha $ and $\beta $ denoting the phases associated
with the Majorana character of neutrinos.

The clear experimental evidence for a non-zero $\theta _{13}$ has motivated
a series of studies on how to generate a non-vanishing $\theta _{13}$
through a small perturbation of the tribimaximal and other schemes which
predict $\theta _{13}=0$ in lowest order. \textbf{The distinctive feature of
our analysis}, is the fact that we start from a non-trivial limit of three
exactly degenerate Majorana neutrinos. In the previous section, we presented
the most general mixing matrix $U_{o}$ in this limit and explained that it
can be parametrized by two angles and one CP violating phase: 
\begin{equation}
U_{o}=O{\small (\theta ,\phi )}\cdot K
\end{equation}%
with $K$ a diagonal matrix such as the one written in Eq.~(\ref{eq8}). This
choice for the matrix $K$ implies that in the CP conserving limit
corresponding to $\alpha =0$ or $\pi $, one neutrino has a CP parity different from
the other two. Otherwise, in the limit of exact degeneracy, with CP
conservation and all neutrinos having the same CP parity, the two angles 
$\phi $ and $\theta $ could be rotated away. 

Lifting the degeneracy corresponds to adding a small perturbation to $S_{o}$ 
\begin{equation}
M={\mu }\ (S_{o}+\varepsilon ^{2}\ Q_{o})  \label{quasi}
\end{equation}%
the matrix $Q_{o}$ is fixed in such a way that the correct neutrino masses
are obtained. It will be a function of the neutrino mass differences given
in terms of $\Delta m_{21}^{2}$ and $\Delta m_{31}^{2}$ defined by:%
\begin{equation}
\begin{array}{l}
\Delta m_{21}^{2}=m_{2}^{2}-m_{1}^{2} \\ 
\Delta m_{31}^{2}=|m_{3}^{2}-m_{1}^{2}|%
\end{array}
\label{m12}
\end{equation}
as well as the overall mass scale ${\mu }$. The parameter 
$\varepsilon ^{2}$ is chosen as: 
\begin{equation}
\varepsilon ^{2}\equiv \frac{\Delta m_{31}^{2}}{2{{\mu }^{2}}}
\label{e}
\end{equation}
Quasidegeneracy forces the overall mass scale to be much larger than the
neutrino mass differences and guarantees the smallness of the perturbation
parameter $\varepsilon ^{2}$. See Table 1 and subsequent comments.\newline

Our strategy for confronting the data on neutrino masses and mixing is the
following: \newline

$(i)$ We assume that the physics responsible for the lifting of the degeneracy,
does not introduce new sources of CP violation beyond the phase $\alpha$, already
present in the limit of exact degeneracy. As a result, after the lifting of the degeneracy, the
leptonic mixing matrix is given by:
\begin{equation}
U_{PMNS}=U_{o}\cdot O  \label{uuo}
\end{equation}
where $O$ is an orthogonal matrix, parametrized by small angles. The fact
that $O$ is orthogonal, rather than a general unitary matrix, implies that $%
U_{PMNS}$ still diagonalizes $S_{o}$, thus establishing a strong connection
between the degenerate and quasidegenerate case. This is particularly
relevant since we shall take as starting point for $U_{o}$ some of the most
interesting examples considered in the literature based on symmetries and
with a zero in the $(13)$ entry of {$U_{o}$}. \newline

$(ii)$ After the lifting of the degeneracy, the single
phase $\alpha $ will generate both Dirac and Majorana-type CP violations.
This is a distinctive feature of our framework. \newline

With the notation of Eq.(\ref{uuo}), $Q_{o}$ introduced in Eq.~(\ref{quasi})
is determined by: 
\begin{equation}
\varepsilon ^{2}\ Q_{o}=U_{o}^{\ast }\cdot O\ \left( \frac{1}{{\mu }
}\ D_{\nu }-{1\>\!\!\!\mathrm{I}}\right) \ O^{T}\cdot U_{o}^{\dagger }\
,\qquad D_{\nu }=diag({m_{\nu }}_{1},{m_{\nu }}_{2},{m_{\nu }}_{3})
\label{q0}
\end{equation}%
In the limit of exact degeneracy the matrix $O$ has no physical meaning, it
only acquires meaning with the lifting of the degeneracy. A striking feature
is the fact that new sources of CP violation are not introduced. However,
once the matrix $O$ is included, the CP violating phase present in $K$
ceases to be a factorizable phase and in general gives rise to Dirac-type CP
violation. \newline

The matrix $O$ will be parametrized by three mixing angles which we denote
by: 
\begin{equation}
O=O_{12}O_{13}O_{23}=\left( 
\begin{array}{ccc}
c_{\phi _{1}} & s_{\phi _{1}} & 0 \\ 
-s_{\phi _{1}} & c_{\phi _{1}} & 0 \\ 
0 & 0 & 1
\end{array}
\right) \left( 
\begin{array}{ccc}
c_{\phi _{3}} & 0 & s_{\phi _{3}} \\ 
0 & 1 & 0 \\ 
-s_{\phi _{3}} & 0 & c_{\phi _{3}}
\end{array}
\right) \left( 
\begin{array}{ccc}
1 & 0 & 0 \\ 
0 & c_{\phi _{2}} & s_{\phi _{2}} \\ 
0 & -s_{\phi _{2}} & c_{\phi _{2}}
\end{array}
\right)  \label{ooo}
\end{equation}
Our choice of $U_{o}$'s is based on the fact that $\theta _{13}$ is known to
be a small angle. Furthermore, in each case, the resulting $O$ matrices
represent small perturbations around $U_{o}$ matrices. Once the matrix $O$
is fixed and the scale ${\mu }$ of neutrino masses is specified, $Q_{o}$ can 
be computed from Eq.~(\ref{q0}) \newline

In our analysis, we use data from the global fit of neutrino oscillations
provided in Ref.~\cite{Tortola:2012te} requiring agreement within $1\sigma $
range. Table 1 summarizes the data obtained from Ref.~\cite{Tortola:2012te}.
From Table 1, assuming ${{\mu }}\sim 0.5$ eV, we obtain 
$\varepsilon ^{2}$ of the order $5\times 10^{-3}$.

In what follows we discuss separately several different cases of interest.

\begin{center}
\begin{table}[h]
\caption{Neutrino oscillation parameter summary. For $\Delta m^2_{31}$, 
$\sin^2 \protect\theta_{23}$ , $\sin^2 \protect\theta_{13}$, and $\protect%
\delta$ the upper (lower) row corresponds to normal (inverted) neutrino mass
hierarchy.}
\label{reps}
\begin{tabular}{ccc}
\hline\hline
Parameter & Best fit & $1 \sigma $ range \\ \hline
$\Delta m^2_{21}$ $[10^{-5} eV^2 ] $ & 7.62 & 7.43 -- 7.81 \\ 
$\Delta m^2_{31}$ $[10^{-3} eV^2 ] $ & 2.55 & 2.46 -- 2.61 \\ 
$\Delta m^2_{31}$ $[10^{-3} eV^2 ] $ & 2.43 & 2.37 -- 2.50 \\ 
$\sin^2 \theta_{12}$ & 0.320 & 0.303 -- 0.336 \\ 
$\sin^2 \theta_{23}$ & 0.613 $(0.427)$ & 0.400 --0. 461 and 0.573 -- 0.635
\\ 
$\sin^2 \theta_{23}$ & 0.600 & 0.569 -- 0.626 \\ 
$\sin^2 \theta_{13}$ & 0.0246 & 0.0218 --0.0275 \\ 
$\sin^2 \theta_{13}$ & 0.0250 & 0.0223 -- 0.0276 \\ 
$\delta$ & 0.80 $\pi $ & 0 --2 $\pi$ \\ 
$\delta$ & -0.03 $\pi$ & 0 --2 $\pi$ \\ \hline
\end{tabular}
\end{table}
\end{center}

\subsection{Perturbing tribimaximal mixing}

In this case our starting point is $U_{o}=U_{TBM}\cdot K$ with: 
\begin{equation}
U_{TBM}=\left( 
\begin{array}{ccc}
\frac{2}{\sqrt{6}} & \frac{1}{\sqrt{3}} & 0 \\ 
\frac{1}{\sqrt{6}} & -\frac{1}{\sqrt{3}} & \frac{1}{\sqrt{2}} \\ 
\frac{1}{\sqrt{6}} & -\frac{1}{\sqrt{3}} & -\frac{1}{\sqrt{2}}%
\end{array}%
\right) \qquad \mbox{and}\qquad K=diag(1,i,e^{-i\alpha /2})
\end{equation}%
In the notation of Eq.~(\ref{eq8}), this ansatz corresponds to $\phi
=45^{\circ }$ and $\cos \left( \frac{\theta }{2}\right) =\frac{2}{\sqrt{6}}$
i.e., $\frac{\theta }{2}=35.26^{\circ }$. We allow the angle $\alpha $ to
vary, together with the three angles of the matrix $O$. In this example,
agreement with the global fit for the experimental values requires lowering
the values for the mixing angles $\theta _{12}$ and $\theta _{23}$ of $U_{o}$
and at the same time generating a $\theta _{13}$ different from zero.

Denoting the entries of $U_{PMNS}$ by $U_{ij}$ we have:%
\begin{equation}
\begin{array}{l}
|U_{11}|=|\frac{2}{\sqrt{6}}O_{11}+\frac{i}{\sqrt{3}}O_{21}|=c_{12}c_{13} \\ 
|U_{12}|=|\frac{2}{\sqrt{6}}O_{12}+\frac{i}{\sqrt{3}}O_{22}|=s_{12}c_{13} \\ 
|U_{13}|=|\frac{2}{\sqrt{6}}O_{13}+\frac{i}{\sqrt{3}}O_{23}|=s_{13} \\ 
|U_{23}|=|\frac{1}{\sqrt{6}}O_{13}-\frac{i}{\sqrt{3}}O_{23}+\frac{1}{\sqrt{2}%
}\ e^{-i\alpha /2}\ O_{33}|=s_{23}c_{13}%
\end{array}
\label{u}
\end{equation}
The first three equations allow to determine $\phi _{1}$, $\phi _{3}$ and $%
\phi _{2}$, the fourth one puts bounds on the phase $\alpha $ thus
constraining the strength of leptonic CP violation \cite{Branco:2011zb}. At
this stage it is worth emphasizing that there is strong experimental
evidence that in the quark sector the $V_{CKM}$ matrix is complex even if
one assumes the possible presence of physics beyond the Standard Model \cite%
{Botella:2005fc}. As a result, it is natural to assume that the leptonic
sector also violates CP.

This scenario allows for a particularly simple solution since, one can reach
agreement with the experimental data by choosing a matrix $O$ with only one
parameter different from zero, namely the angle $\phi _{2}$. In this case
the relevant $O_{ij}$ simplify significantly and one can express $\sin
^{2}\left( \theta _{12}\right) $, $\sin ^{2}\left( \theta _{23}\right) $ and 
$\sin ^{2}\left( \theta _{13}\right) $ simply in terms of $\phi _{2}$, and
the phase $\alpha $, or else. equivalently, in terms of $|U_{13}|$ and the
phase $\alpha $:

\begin{eqnarray}
&&\sin ^{2}\left( \theta _{13}\right) \equiv |U_{13}|^{2}=\frac{\sin
^{2}\left( \phi _{2}\right) }{3}  \label{u13} \\
&&\sin ^{2}\left( \theta _{12}\right) \equiv \sin ^{2}(\theta _{solar})=%
\frac{1-\sin ^{2}\phi _{2}}{3-\sin ^{2}\phi _{2}}=\frac{\frac{1}{3}%
-|U_{13}|^{2}}{1-|U_{13}|^{2}}  \label{sol} \\
&&\sin ^{2}\left( \theta _{23}\right) \equiv \sin ^{2}(\theta _{atm})=\frac{1%
}{2}-\frac{\sqrt{6}\sin (\frac{\alpha }{2})\sin \phi _{2}\cos \phi _{2}}{%
3-\sin ^{2}\phi _{2}}=\frac{1}{2}-\frac{\sqrt{2}\sin (\frac{\alpha }{2})%
{\small |U}_{13}{\small |}\sqrt{1-3{\small |U}_{13}{\small |}^{2}}}{1-%
{\small |U}_{13}{\small |}^{2}}  \label{atm}
\end{eqnarray}
Clearly, $|U_{13}|$ fixes the allowed range for the angle $\phi _{2}$ and in
this limit only $\sin ^{2}(\theta _{atm})$ depends on the phase $\alpha $.
From Eq.~(\ref{u13}) and taking the best fit value from Table 1 we obtain $%
\sin \left( \phi _{2}\right) =0.27$. It is instructive to determine $Q_{o}$
for this value of $\sin \left( \phi _{2}\right) $. Making use of Eq.~(\ref%
{q0}), and keeping only the dominant terms, by making the following
approximations: 
\begin{equation}
{\mu }\sqrt{1+\frac{\Delta m_{21}^{2}}{{\mu }^{2}}}\simeq 
{\mu }\qquad ;\qquad {\mu }\sqrt{1+\frac{\Delta m_{31}^{2}%
}{{{\mu }^{2}}}}\simeq {\mu }\left( 1+\frac{\Delta
m_{31}^{2}}{2{\mu }^{2}}\right) ={\mu }(1+\varepsilon ^{2})
\label{mu}
\end{equation}
the explicit expression for $Q_{o}$ simplifies to: 
\begin{equation}
Q_{o}=\left( 
\begin{array}{lll}
-0.0243 & 0.0243-0.1061e^{i\frac{\alpha }{2}} & 0.0243+0.1061e^{i\frac{
\alpha }{2}} \\ 
0.0243-0.1061e^{i\frac{\alpha }{2}} & (0.1559i+0.6808e^{i\frac{\alpha }{2}
})^{2} & -0.0243-0.4634e^{i\alpha } \\ 
0.0243+0.1061e^{i\frac{\alpha }{2}} & -0.0243-0.4634e^{i\alpha } & 
(0.1559i-0.6808e^{i\frac{\alpha }{2}})^{2}
\end{array}
\right)   \label{qo}
\end{equation}
It should be noticed that, even after factoring out $\varepsilon ^{2}$, most
entries of the matrix $Q_{o}$ have modulus much smaller than one, thus confirming
that we are doing a very small perturbation around the degeneracy limit.

We find that the angle $\phi _{2}$ cannot deviate significantly from the
value of the Cabibbo angle. The constraints on the phase $\alpha $ obtained
from Eq.~(\ref{atm}), translate into bounds for the Dirac CP violating phase 
$\delta $. The strength of Dirac-type CP violation is often given in terms
of the modulus of the parameter $I_{CP}$ defined as the imaginary part of a
quartet of the mixing matrix $U_{PMNS}$, i.e., $I_{CP}\equiv \mbox{Im}
|U_{ij}U_{ik}^{\ast }U_{lj}^{\ast }U_{lk}|$ with $i\neq l$, $j\neq k$. Due
to the unitariry of $U_{PMNS}$ all quartets have the same modulus. For the
standard parametrization, given in Eq.~(\ref{pdg}), we have: 
\begin{equation}
I_{CP}\equiv \frac{1}{8}\left\vert \sin (2\theta _{12})\sin (2\theta
_{13})\sin (2\theta _{23})\cos (\theta _{13})\sin \left( \delta \right)
\right\vert
\end{equation}
In our framework, with only $\phi _{2}$ and $\alpha $ different from zero, 
$I_{CP}$ is given by: 
\begin{equation}
I_{CP}=\left\vert \frac{\cos (\alpha /2)\sin \left( \phi _{2}\right) \cos
\left( \phi _{2}\right) }{3\sqrt{6}}\right\vert
\end{equation}%
and is predicted to be of order $10^{-2}$, meaning that it could be within
reach of future neutrino experiments. This is a special prediction for this
framework since from the values of Table 1 we can conclude that the
experimental bounds at $1\sigma $ level allow for the leptonic strength of
Dirac-type CP violation to range from 0 to about $4\times 10^{-2}$. 
\begin{figure}[h]
\begin{center}
\includegraphics[scale=0.8]{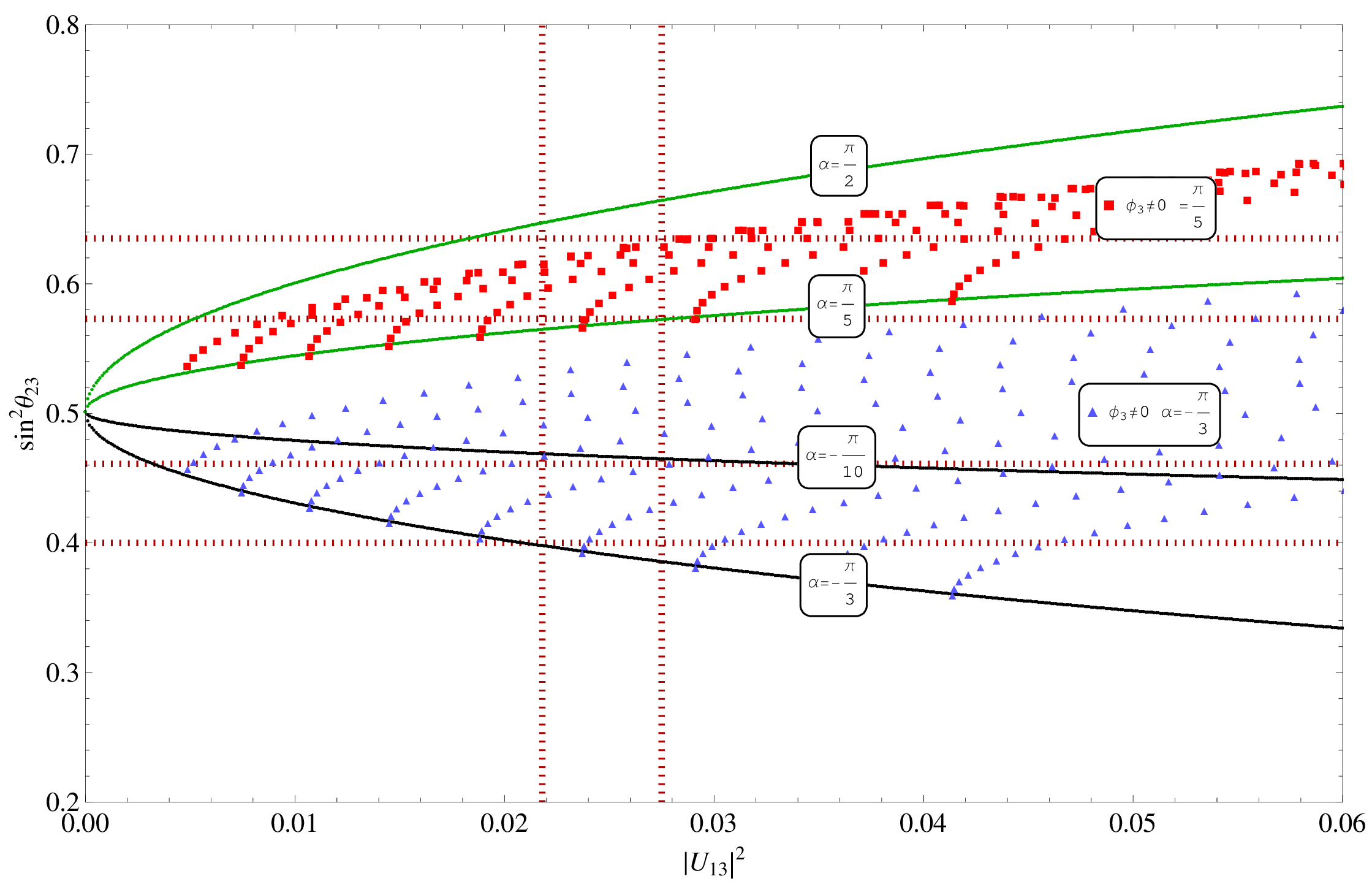}
\end{center}
\caption{$\sin ^{2}\protect\theta _{23}$ versus $|U_{13}|^{2}$ obtained by
perturbing tribimaximal mixing with $\protect\phi _{3}=0$. Each curve
corresponds to a fixed $\protect\alpha $ and to $\protect\phi _{1}=0$ ,
therefore $\protect\phi _{2}$ is the only variable. The points drifting away
from each curve were obtained by varying also $\protect\phi _{3}$.}
\label{figura2}
\end{figure}
In Figure 2 we present $\sin ^{2}(\theta _{atm})$ versus $|U_{13}|^{2}$. The
dotted vertical lines delimit the allowed experimental values for 
$|U_{13}|^{2}$. The dotted horizontal lines delimit the two allowed
experimental regions for $\sin ^{2}(\theta _{atm})$ according to Table 1.
The authors of Ref.~\cite{Tortola:2012te} consider the region of lower $\sin
^{2}(\theta _{atm})$ to be experimentally favoured, therefore in our
analysis we require that this region can be reached even though we also
indicate the above region. The different solid lines correspond to our
framework with only one parameter different from zero, the angle $\phi _{2}$,
and for different values of the phase $\alpha $ as indicated in the
figure. The values for this phase are chosen in such a way as to give an
indication of the intervals that are compatible with the experimental data.
Points represented by squares and triangles where obtained with one
additional mixing angle, $\phi _{3}$, different from zero. Squares and
triangles correspond to different values of the phase $\alpha $
respectively, as indicated in the figure. 
\begin{figure}[h]
\begin{center}
\includegraphics[scale=0.8]{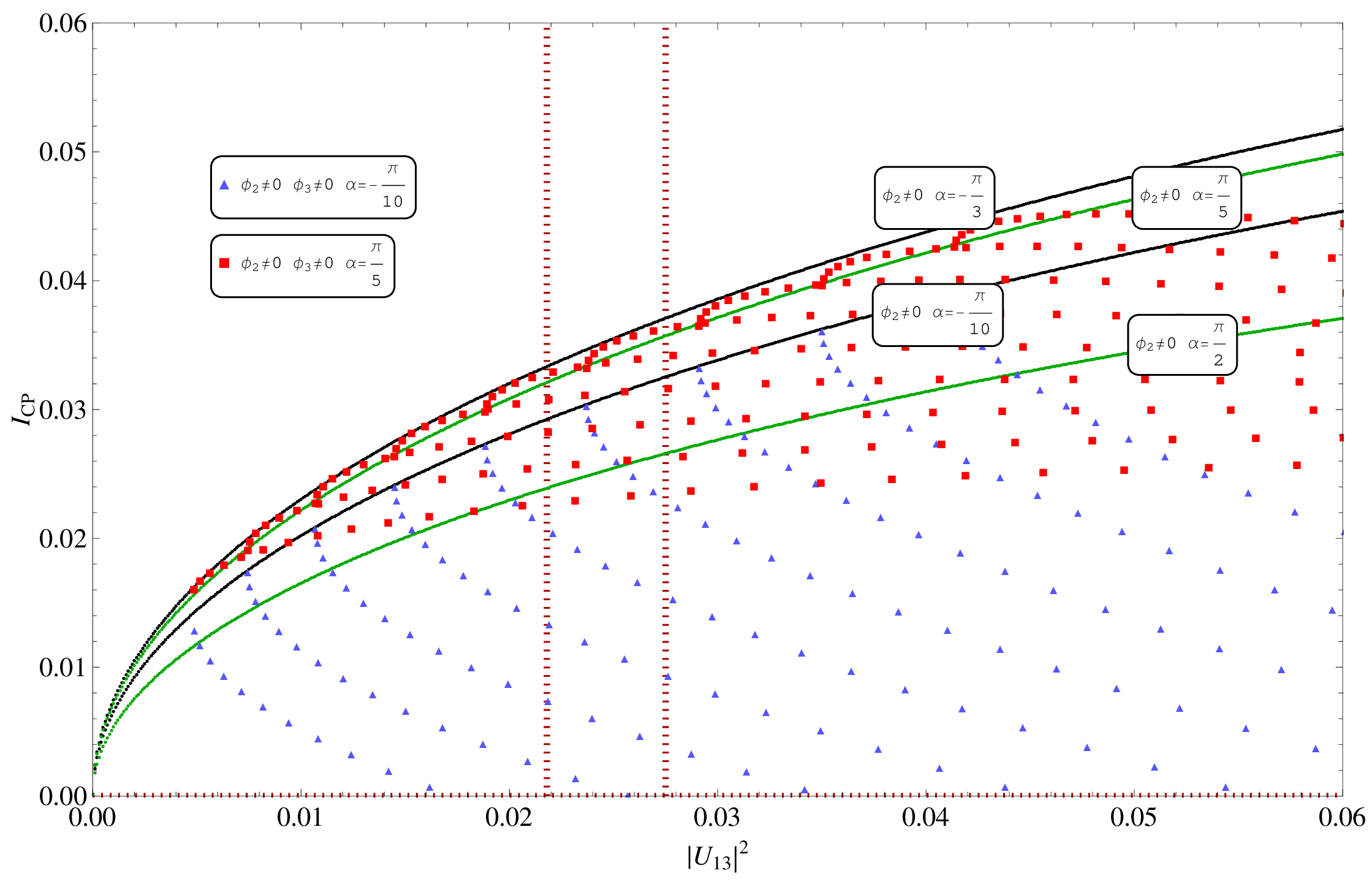}
\end{center}
\caption{$I_{CP}$ versus $|U_{13}|^{2}$ obtained by perturbing tribimaximal
mixing with $\protect\phi _{3}=0$. Each curve corresponds to a fixed $%
\protect\alpha $ and to $\protect\phi _{1}=0$ , therefore $\protect\phi _{2}$
is the only variable. The points drifting away from each curve were obtained
by varying also $\protect\phi _{3}$.}
\label{figura3}
\end{figure}
In Figure 3 we plot $I_{CP}$ versus $|U_{13}|^{2}$. Again the dotted
vertical lines delimit the allowed experimental values for $|U_{13}|^{2}$
and the different solid lines correspond to our framework with only one
mixing angle different from zero and for different values of the phase $%
\alpha $ as indicated in the figure. The values chosen for this phase are
based on the information contained in Figure 2. Points represented by
squares and triangles where obtained with one additional mixing angle
different from zero, which in this case was chosen to be $\phi _{1}$.
Squares and triangles correspond to different values of the phase $\alpha $
respectively, as indicated in the figure.

Concerning the neutrinoless double beta decay, this process depends on the
effective Majorana mass, $m_{ee}$, defined by; 
\begin{equation}
m_{ee}=\left\vert \sum_{k=1}^{3}U_{1k}^{2}m_{k}\right\vert 
\end{equation}%
In the above framework, the dominant terms, ignoring in particular
corrections of order $\Delta m_{21}^{2}/{\mu }^{2}$ are:
\begin{equation}
\begin{array}{l}
m_{ee}=\frac{{\mu }\ }{3}\ \left( 1-\frac{\Delta m_{31}^{2}}{2
{\mu }^{2}}\sin ^{2}(\phi _{2})\right)  \\ 
\\ 
=\frac{{\mu }}{3}\ \ \left( 1-3\frac{\Delta m_{31}^{2}}{2{%
\mu }^{2}}|U_{13}|^{2}\right) 
\end{array}
\label{mee}
\end{equation}
Agreement with the present experimental bounds, \cite{mee}
taking into account nuclear
physics uncertainties \cite{Vergados:2012xy} requires $|m_{ee}|$ to be
smaller than $0.4$ $eV$. The Heidelberg-Moscow experiment 
\cite{KlapdorKleingrothaus:2000sn} claimed to have obtained a non-zero result
close to $0.38$ $eV$ which would imply all three neutrino masses close to $1$
$eV$. These masses are somewhat above the bound favoured by cosmology,
however the cosmological bound depends on model assumptions and on the data
set that is taken into consideration \cite{Wong:2011ip}.

In this framework, the angle $\phi _{2}$ cannot deviate significantly from
the value of the Cabibbo angle even when we extend it to include other
non-zero mixing angles. In fact, the range of the allowed experimental
parameters given in Table 1 can accommodate non zero values for the two
other angles in the matrix $O$ requiring them to be smaller than the Cabibbo
angle. In this case the simple expressions given above must be replaced by
somewhat more cumbersome and less transparent ones. The solar angle obtained
in the unperturbed tribimaximal mixing case is larger than the allowed experimental
values. The angle $\phi _{2}$ is the only one in $O$ capable of lowering its
value. The effect of the other two mixing angles is the opposite.

\subsection{Perturbing other interesting schemes}

As stated before, we analysed perturbations around some of the well known
mixing textures considered in the literature with a zero in the (13) entry.
Examples of such textures include the democratic mixing, $U_{DM}$ 
\cite{Fritzsch:1995dj}, bimaximal mixing $U_{BM}$ \cite{Barger:1998ta}, golden
ratio mixings $U_{GRM1}$ \cite{Kajiyama:2007gx}, $U_{GRM2}$ 
\cite{Rodejohann:2008ir}, hexagonal mixing $U_{HM}$ \cite{Albright:2010ap} and
bidodeca mixing $U_{BDM}$ \cite{Kim:2010zub}, \cite{Kim:2011vr}. The
democratic mixing and the bimaximal mixing are of the form: 
\begin{equation*}
U_{DM}=\left( 
\begin{array}{ccc}
\frac{1}{\sqrt{2}} & \frac{1}{\sqrt{2}} & 0 \\ 
-\frac{1}{\sqrt{6}} & \frac{1}{\sqrt{6}} & \sqrt{\frac{2}{3}} \\ 
\frac{1}{\sqrt{3}} & -\frac{1}{\sqrt{3}} & \frac{1}{\sqrt{3}}
\end{array}
\right) ;\qquad U_{BM}=\left( 
\begin{array}{ccc}
\frac{1}{\sqrt{2}} & \frac{1}{\sqrt{2}} & 0 \\ 
-\frac{1}{2} & \frac{1}{2} & \frac{1}{\sqrt{2}} \\ 
\frac{1}{2} & -\frac{1}{2} & \frac{1}{\sqrt{2}}
\end{array}
\right)
\end{equation*}
These two cases are very constrained in our framework, since they correspond
to $\sin ^{2}(\theta _{sol})=0.5$ which lies significantly above the
favoured experimental range given in Table 1. Although it is still possible
to bring it down to acceptable values making use of $\phi _{2}$, agreement
with the experimental values given in Table 1 is hardly possible at 1$\sigma$
level. Therefore, we do not further analyse these two cases.

The other textures mentioned above are $U_{GRM1}$, $U_{GRM2}$ and the
hexagonal mixing $U_{HM}$ which coincides with the bidodeca mixing $U_{BDM}$:

\begin{equation*}
{\tiny U_{GRM1}=\left( 
\begin{array}{ccc}
\sqrt{\frac{1}{2}\left( 1+\frac{1}{\sqrt{5}}\right) } & \sqrt{\frac{2}{5+%
\sqrt{5}}} & 0 \\ 
-\frac{1}{\sqrt{5+\sqrt{5}}} & \frac{1}{2}\sqrt{1+\frac{1}{\sqrt{5}}} & 
\frac{1}{\sqrt{2}} \\ 
\frac{1}{\sqrt{5+\sqrt{5}}} & -\frac{1}{2}\sqrt{1+\frac{1}{\sqrt{5}}} & 
\frac{1}{\sqrt{2}}%
\end{array}
\right) ;\qquad U_{GRM2}=\left( 
\begin{array}{ccc}
\frac{1}{4}(1+\sqrt{5}) & \frac{1}{2}\sqrt{\frac{1}{2}(5-\sqrt{5})} & 0 \\ 
-\frac{1}{4}\sqrt{5-\sqrt{5}} & \frac{1+\sqrt{5}}{4\sqrt{2}} & \frac{1}
{\sqrt{2}} \\ 
\frac{1}{4}\sqrt{5-\sqrt{5}} & -\frac{1+\sqrt{5}}{4\sqrt{2}} & \frac{1}
{\sqrt{2}}%
\end{array}
\right) }
\end{equation*}
\begin{equation*}
U_{HM}=U_{BDM}=\left( 
\begin{array}{ccc}
\frac{\sqrt{3}}{2} & \frac{1}{2} & 0 \\ 
-\frac{1}{2\sqrt{2}} & \frac{1}{2}\sqrt{\frac{3}{2}} & \frac{1}{\sqrt{2}} \\ 
\frac{1}{2\sqrt{2}} & -\frac{1}{2}\sqrt{\frac{3}{2}} & \frac{1}{\sqrt{2}}
\end{array}
\right)
\end{equation*}
The case of the golden ratio mixing 2 is less favourable than the golden
ratio mixing 1, due to the fact that the corresponding solar angle is
larger. We analysed in more detail only the cases starting with $U_{GRM1}$
and $U_{HM}$ . We have scanned the allowed region of parameter space for the
angles $\phi _{1}$, $\phi _{3}$, $\phi _{2}$ of our perturbation, and for
the phase $\alpha $. Both examples have very similar features. The exact
analytic expressions are obtained from Eqs.~(\ref{uuo}) and (\ref{ooo}). A
novel feature of these examples is the fact that agreement with experiment
cannot be obtained with the matrix $O$ parametrized by one mixing angle
only. Furthermore, unlike the tribimaximal mixing case, it is $\phi _{1}$
that is required to differ from zero and on the other hand either $\phi _{2}$
or $\phi _{3}$ can be zero, although not simultaneously. These new features
are related to the fact that in both cases the corresponding solar angle
lies below the experimental range unlike in the tribimaximal case. As
pointed out, in the tribimaximal case the angle $\phi _{2}$ played a
fundamental r\^{o}le in lowering this angle. In the case of $\phi _{3}$
equal to zero, $U_{13}$ is then given by: 
\begin{equation}
U_{13}=({U_{o}})_{11}\sin \left( \phi _{1}\right) \sin \left( \phi
_{2}\right) +({U_{o}})_{12}\cos \left( \phi _{1}\right) \sin \left( \phi
_{2}\right)  \label{us}
\end{equation}%
it is the second term that gives the dominant contribution. The fact that
there are two independent parameters in the matrix $O$ does not allow to
express $\sin ^{2}(\theta _{solar})$, $\sin ^{2}(\theta _{atm})$ and $I_{CP}$
in terms of $|U_{13}|$ only. 
\begin{figure}[t]
\begin{center}
\includegraphics[scale=0.65]{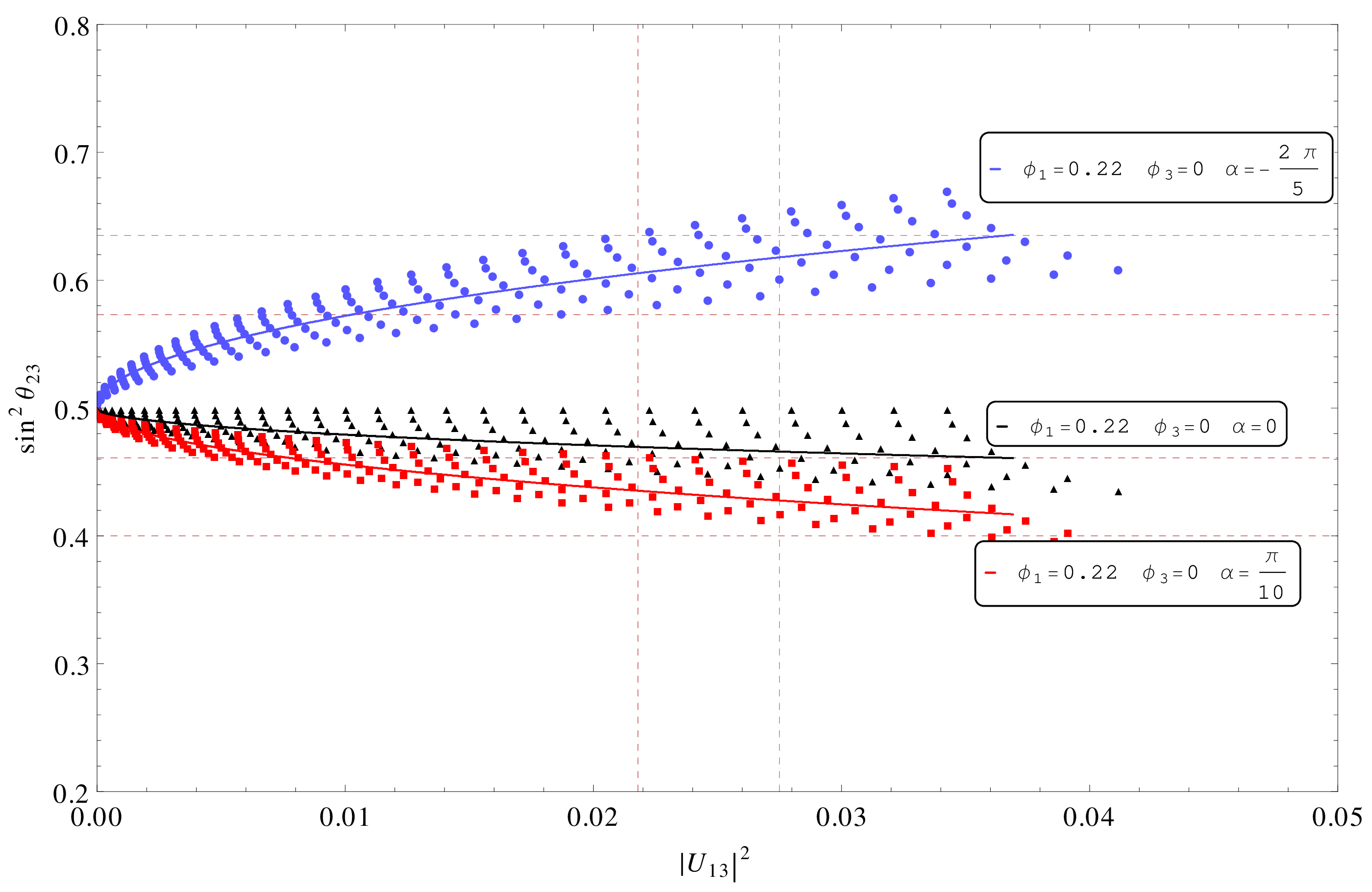}
\end{center}
\caption{$\sin ^{2}\protect\theta _{23}$ versus $|U_{13}|^{2}$ obtained by
perturbing golden ratio 1 with $\protect\phi _{3}=0$. Each curve corresponds
to a fixed $\protect\alpha $ and fixed value of $\protect\phi _{1}$. The
points drifting away from each curve were obtained by varying $\protect\phi %
_{1}$.}
\label{figura4}
\end{figure}
However it is still instructive to plot these quantities as a function of $%
|U_{13}|$ for certain choices of the parameters of the matrix $O$. For
illustration, we present in Figure 4 $\sin ^{2}\left( \theta _{23}\right) $
versus $|U_{13}|^{2}$ with $\sin \left( \phi _{3}\right) =0$. This plot is
done for golden ratio 1. The hexagonal mixing case presents similar
features. Each curve in the figure corresponds to a fixed value of the
parameter $\phi _{1}$ and of the phase $\alpha $ and is therefore obtained
by varying $\phi _{2}$. The points drifting away from each curve were
obtained by varying in turn $\phi _{1}$ still keeping $\alpha $ fixed and $%
\phi _{3}=0$ . Circles, triangles and squares are associated to different
choices of the phase $\alpha $, respectively, as indicated in the figure.
The figure shows that it is possible to accommodate $\alpha =0$,
corresponding to the CP conserving case, however agreement with experiment
in this case is only possible for a small range of the parameter space. On
the other hand, fixing $\phi _{3}=0$ allows both $\phi _{1}$ and $\phi _{2}$
to be close to the Cabibbo angle.

\section{Conclusions}

In this paper, we present a novel proposal for the understanding of the
observed pattern of leptonic mixing, which relies on the assumption that
neutrinos are Majorana particles. It is argued that the observed large
leptonic mixing may arise from a quasidegeneracy of three Majorana
neutrinos. The essential point is the fact that the limit of exact mass
degeneracy of three Majorana neutrinos is non-trivial as lepton mixing and
even CP violation can arise. This limit is particularly interesting since in
this case leptonic mixing can be parametrized by only two mixing angles and one
phase, implying that without loss of generality the leptonic mixing matrix
can be written with one zero entry.  We have then conjectured that the 
smallness of $|U_{13}|$ when compared to the other elements of $U_{PMNS}$ 
may result from this fact. We  show that the observed pattern of mixing and 
neutrino mass differences can be generated through a small perturbation 
of the exact degenerate case, without the introduction of additional CP violating 
sources. A key point in our work is the assumption that the physics responsible
for the lifting of the degeneracy does not introduce new sources of CP violation.
Our perturbation requires the multiplication on the right by an orthogonal 
matrix. The resulting unitary matrix $U_{o}O$ which can be identified as
the $U_{PMNS}$ matrix, also diagonalizes the neutrino mass matrix in the fully
degenerate case.
This allows to establish a strong  connection between the degenerate and 
quasidegenerate cases and at the same time reducing the number of free parameters.
Upon the lifting of degeneracy, this single phase generates both Majorana
and Dirac-type CP violation in the leptonic sector. For definiteness, we
have used as the starting point for the perturbation around the limit of
exact degeneracy, some of the most interesting Ans\"{a}tze considered in the
literature, which were proposed in the past assuming $\theta_{13} =0$.
We analyse correlations among
physical observables, and point out that in most of the cases
considered, the implied strength of leptonic Dirac-type CP violation is
large enough to be detectable in the next round of experiments.

\section*{Acknowledgments}

This work is partially supported by Funda\c{c}\~ao para a Ci\^encia e a
Tecnologia (FCT, Portugal) through the projects CERN/FP/123580/2011,
PTDC/FIS-NUC/0548/2012, EXPL/FIS-NUC/0460/2013, and CFTP-FCT Unit 777
(PEst-OE/FIS/UI0777/2013) which are partially funded through POCTI (FEDER),
COMPETE, QREN and EU. DW is presently supported by a postdoctoral fellowship
of the project CERN/FP/123580/2011 and his work is done at CFTP-FCT Unit 777.

\end{document}